\documentclass[twocolumn,showpacs,preprintnumbers,amsmath,amssymb,floatfix]{revtex4}
\usepackage{amsmath,graphicx,amsfonts}
\usepackage{dcolumn}

 \newcommand{\ket}[1]{| #1 \rangle}
 \newcommand{\bra}[1]{\langle #1 |}
 
\begin{document}

\title{Quantum signatures of breather-breather interactions}

\author{J. Dorignac}
\email{J.Dorignac@hw.ac.uk}

\author{J. C. Eilbeck}
\email{J.C.Eilbeck@hw.ac.uk}

\affiliation{Department of Mathematics, Heriot-Watt University,
  Riccarton, Edinburgh, EH14 4AS, UK }

\author{M. Salerno}
\email{salerno@sa.infn.it} 
\affiliation{Dipartimento di Scienze Fisiche ``E.~R.~Caianiello'' and
  INFM Unit\'a di Salerno, via S.Allende, I-84081, Baronissi (SA),
  Italy}

\author{A. C. Scott}
\email{rover@theriver.com}
\affiliation{Department of Mathematics, University of Arizona, Tucson,
AZ 85721, USA}

\date{\today}

\begin{abstract}
  The spectrum of the Quantum Discrete Nonlinear Schr\"odinger
  equation on a periodic 1D lattice shows some interesting detailed
  band structure which may be interpreted as the quantum signature of
  a two-breather interaction in the classical case.  We show that this
  fine structure can be interpreted using degenerate perturbation
  theory.

\end{abstract}

\pacs{63.20.Pw}
\keywords{Anharmonic quantum lattices, Quantum breathers, Quantum
  lattice solitons}

\maketitle
\section{Introduction}
The localization and transport of energy in lattices by intrinsic
localized modes or discrete breathers, has recently been the subject
of intense theoretical and experimental investigation ( see
\cite{vmz03} and references therein).  Corresponding results on the
quantum equivalent of these states are less numerous, c.f.~
\cite{seg94,mk00%,fl03
} for some theoretical results and
\cite{fck98%,sw99,ani00,sms02
} for some experimental work.  Studies of
quantum modes on small lattices are of increasing interest for quantum
devices based on quantum dots (c.f.\ \cite{li03}) and for studies of
photonic crystals.

We present some results on breather bands in small one dimensional
lattices with a small number of quanta. We study a periodic lattice
with $f$ sites containing bosons, described by the quantum version of
the discrete nonlinear Schr\"{o}dinger equation (DNLS) which
will be denoted by QDNLS. It has the Hamiltonian
\begin{equation} \label{Ham1}
H_1 = -\sum_{s=1}^f\left[\gamma_1 a_s^{\dagger} a_s^{\dagger}a_s a_s  
  +\epsilon a_{s+1}^{\dagger}a_s+ \text{h.c.}\right],
\end{equation}
which conserves the number of quanta $n$. This model, also known as
the Bose-Hubbard model, is presently used to investigate cold bosonic
atoms in optical lattices \cite{Jaksch98}.

Assuming that the anharmonic parameter $\gamma_1>0$ is stronger than the
intersite coupling, the eigenvalues of (\ref{Ham1}), plotted as a
function of the wave number $k$, separate out into distinct bands
\cite{seg94}.  The eigenstates of each band are dominated by states in
which the bosons have clumped together, two or more on one site.  For
example, with $n=4$, the lowest band is, to a good approximation, a
linear combination of states with 4 bosons on site $i$ and no bosons
elsewhere.  The next lowest band is mostly composed of states with 3
quanta on one site and another quanta elsewhere.  The third band is
mostly composed of states with 2 quanta on one site and 2 quanta on a
separate site.  We will refer to this band as a $\{2,2\}$ band in an
obvious notation.  This band is of great interest since it represents the
simplest case of a band describing two ``composite'' particles interacting
with each other.  Our letter is devoted to the fine structure of this
and similar bands such as the $\{4,2\}$ and $\{3,3\}$ bands in the
$n=6$ case.

\begin{figure}
\includegraphics[scale=0.49]{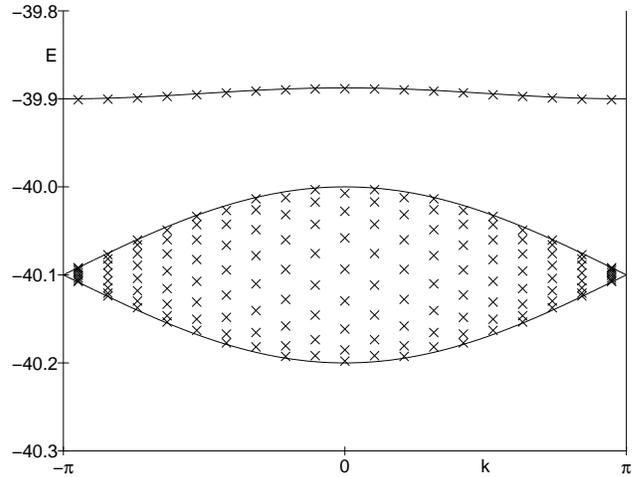}
\caption{\label{fig1} Detail of the eigenvalue spectrum for the 
  QDNLS periodic lattice (\ref{Ham1}), $n=4, f=19, \gamma_1=10,
  \epsilon=0.5$.}
\end{figure}

The fine structure of the $\{2,2\}$ band, see Fig.~\ref{fig1},
shows the eigenvalues (crosses) in the $n=4$ case. 
The solid lines show the results of the theoretical calculations
(described below) in the asymptotic limit $f \rightarrow \infty$.  We
stress again that this picture shows only {\em one} of the bands in
the $n=4$ spectrum.  Its fine detail is revealed as a ``continuum''
band (in the $f \rightarrow \infty$ limit), plus a single $k$-dependent
``line band''.  Examination of the corresponding eigenvectors shows
that the ``line band'' is mostly composed of states where the two
sites each with two quanta lie occupy {\em adjacent} sites, whereas in
the ``continuum'' band these sites are separated by one or more vacant
sites.
  
\section{Generalized Hamiltonian}
The standard QDNLS Hamiltonian is 
thermodynamically unstable, i.e. the energy of the ground state goes
to $-\infty$ as the number of particles goes to $+\infty$.  Moreover, 
some bands can be degenerate with others, for example when $n=6$,
the $\{3,3\}$ band overlaps with the $\{4,1,1\}$ band.  In some cases
it would be convenient for the bands under study to be the lowest
energy bands of the system.

For these reasons we consider also the following generalized QDNLS
Hamiltonian
\begin{equation} \label{Ham2}
H_2 = H_1 +\gamma_2 \sum_{s=1}^f
  a_s^{\dagger}a_s^{\dagger}a_s^{\dagger} a_s a_s a_s \, . 
\end{equation}
The $\gamma_2>0$ term is a saturation term which discourages too many
bosons occupying the same site. A similar term appears also in
nonlinear optics \cite{mcgsq02} and in cold bosonic atoms trapped in
optical lattices \cite{gammal00} where it describes three body interactions.
By varying the values of $\gamma_1$ and
$\gamma_2$ it is possible to ``shuffle'' the order of bands in the
$(E,k)$ diagram. 

At zero coupling ($\epsilon=0$), we denote by $E_s$ the respective
energies of $s$ bosons on the same site, $ E_1 = 0, \quad E_2 =
-2\gamma_1, \quad E_3 = -6 (\gamma_1-\gamma_2), \quad E_4 = -12
(\gamma_1-2\gamma_2)$, etc.  Thus in the $n=4$ case the $\{4\}$ band
has energy $E_4$, the $\{31\}$ band has energy $E_3$, the $\{22\}$
band energy $2E_2$, etc. Providing $\gamma_1<3\gamma_2$, the bottom of
the $\{22\}$ band is the {\em ground state} of the system.

\section{Perturbation Theory}

%\subsection{Notation}
To describe the components of the quantum states we use a position
state representation, where for example the state $\ket{020020\dots}$
represents a state with two quanta at the lattice position 2, two
quanta at the lattice position 5, and zero quanta elsewhere.  In view
of the periodic nature of the lattice, we can generate an equivalence
class of states by applying the translation operator (with periodic
boundary condition) to one of these states.  We will refer to these
classes by ordering them such that the leftmost number is the largest,
so for example $\ket{3000\dots}$ is shorthand for $\ket{3000\dots},
\ket{0300\dots}, \ket{0030\dots}$, etc.  For further conciseness we
will truncate all trailing zeros, so the above class becomes $\ket{3}$.
The set of all the classes containing  $\ket{22}, \ket{202},
\ket{2002}$, etc. is referred to as the $\{2,2\}$ band.

At zero coupling, all the $\{2,2\}$ states $\ket{22}$, $\ket{202},
\ket{2002}, \dots $ are degenerate. We use degenerate
perturbation theory to obtain both eigenvalues and eigenstates for
$\epsilon \neq 0$.  For the sake of simplicity, we consider an odd
number of sites $f=2\sigma+1$. 
Bloch waves of $\{2,2\}$ states can be written (in 
the notation of \cite{sc99})
\begin{equation} \label{Blw}
|\psi\rangle  = \sum_{j=1}^{\sigma} c_j \ket{\psi_j}
\end{equation}
where
\begin{equation}\label{Blw1}
\ket{\psi_j} = \frac{1}{\sqrt{f}} \sum_{s=1}^{f} (\hat
T/\tau)^{s-1} \ket{2,\underbrace{0,\dots,0}_{j-1},2}  
\end{equation}
Here $\hat T$ is the translation operator,
$\tau =e^{ik}$, and the crystal momentum 
$k=2\pi l/f$ where $l \in \{-\sigma,\dots,\sigma\}$.

Using (\ref{Blw}) and standard Brillouin-Wigner perturbation theory up
to second order in $\epsilon$ we obtain
\begin{align} \label{Eigeq}
(E - 2E_2) c_{j} &= \sum_{j'=1}^{\sigma}
\bra{\psi_j}V \ket{\psi_{j'}} c_{j'} +\\
&\quad+ \sum_{j'=1}^{\sigma}\sum_{\tilde \psi}\frac{\bra{\psi_j}
  V\ket{\tilde  \psi}
\bra{\tilde \psi}  V\ket{\psi_{j'}}}
{2E_2-\tilde E} c_{j'}  \nonumber 
\end{align}
where $V$ is the hopping term in the Hamiltonian, and $\ket{\tilde
  \psi}$ is any state not in the $\{2,2\}$ subspace.
%but connected to a
%$\{2,2\}$ state by one boson hopping to a nearest neighbor site.
$\tilde E$ is the energy corresponding to $\ket{\tilde \psi}$ in the
uncoupled limit ($\epsilon=0$).

It is obvious that the first sum of (\ref{Eigeq}) is zero as $V$ does
not link any of the $\ket{\psi_s}$ to each other.  In evaluating the
second sum, the $\bra{\cdot}V\ket{\cdot}$ terms are mostly zero except
for the following
\begin{align*}
&\bra{22}V\ket{112}=e^{ik}\surd2\epsilon=e^{ik} \bra{22}V\ket{211},\\
&\bra{22}V\ket{13}=\surd6\epsilon=\bra{22}V\ket{31},\\
&\bra{202}V\ket{112}=\surd2\epsilon=\bra{202}V\ket{211},\\
&\bra{202}V\ket{1102}=e^{ik}\surd2\epsilon=e^{ik}\bra{202}V\ket{2011},\,
\dots
\nonumber 
\end{align*}
We now define the matrix $H^{(22)}_{j,j'}$ by 
\begin{equation}
H^{(22)}_{j,j'}= \sum_{\tilde \psi} \frac{\bra{\psi_j}
  V\ket{\tilde  \psi}
\bra{\tilde \psi}  V\ket{\psi_{j'}}}
{2E_2-\tilde E} \, .
\end{equation}
Then
\begin{equation} \label{H2}
H^{(22)} = \frac{8\epsilon^2}{E_2-2E_1} \, I_{\sigma} +
\frac{4\epsilon^2}{E_2-2E_1} \left( 
\begin{array}{ccccc}
\Gamma & \kappa^* & & &  \\
\kappa & 0 & \kappa^* & & \\
& \ddots & \ddots & \ddots & \\
& & \kappa & 0 & \kappa^*  \\
& & & \kappa & p
\end{array}
\right)
\end{equation}   
where $I_m$ is the $m\times m$ unity matrix and 
\begin{eqnarray} 
\Gamma &=&
\frac{3(E_2-2E_1)}
{2E_2-E_3-E_1}-1 = \frac{3\gamma_2-4\gamma_1}
{\gamma_1-3\gamma_2} \, , \label{Gamma}\\
\kappa &=& e^{i\frac{k}{2}} \cos \left( \frac{k}{2}\right) \
\text{and}\  p = \cos \sigma k \, .   
\end{eqnarray} 

The structure of the matrix (\ref{H2}) is very similar to the
two-quanta case described in \cite{seg94}.  The first term represents
a global shift of the $\{2,2\}$-band, whereas the ``impurity''
$\Gamma$ will in general be responsible for a splitting of the states
$\ket{22}$ from the rest of the band ($\ket{202}$, $\ket{2002}$,
etc.). In this respect, the $\ket{22}$'s can be seen as ``bound states
of doublets'' within the $\{2,2\}$-band.  It is separated from the
continuum band because in addition to linking with states such as
$\ket{211}$, it also links to states such as $\ket{31}$.  This
explains the $\{2,2\}$-band fine structure.

In case the number of sites tends to infinity, (\ref{H2}) may be
diagonalised exactly to yield
\begin{equation} \label{infty}
E = 2E_2 +\left\{\! \begin{array}{c}
\frac{4\epsilon^2}{E_2-2E_1}\left(2+\Gamma +
  \frac{\cos^2(k/2)}{\Gamma} \right) \ \text{iff} \ |\Gamma| >  \cos
\left(\frac{k}{2}\right) \\
\\
\frac{8\epsilon^2}{E_2-2E_1}\left(1+ \cos
  \left(\frac{k}{2}\right) \cos \theta \right)\ \text{where}\ \theta
\in (0,\pi) 
\end{array} \right. 
\end{equation} 

Depending on the various values of $\gamma_1, \gamma_2$, and $\epsilon$,
we can get the line band completely above or below the continuum band,
or partially merged with the continuum band.
\begin{figure}
\includegraphics[scale=0.49]{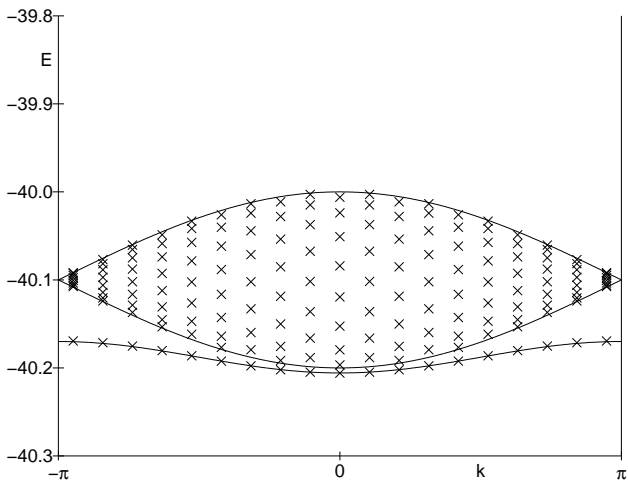}
\caption{\label{fig2} Detail of the eigenvalue spectrum for the generalized
  QDNLS periodic lattice (\ref{Ham2}), $n=4, f=19, \gamma_1=10, \gamma_2=7.5,
  \epsilon=0.5$.}
\end{figure}
Fig \ref{fig2}
shows a case where the line part is below the continuum part of the $\{2,2\}$
band and represents the ground state of the $n=4$ sector.

To provide some insight in the way the two groups of two quanta
interact within a $\ket{22}$ state, we compare its effective mass
$m_{22}^*$ with twice the effective mass $m_{2}^*$ of a single
$\ket{2}$ state. From \cite{seg94} and \eqref{infty}, we obtain
$m_{2}^*\simeq \gamma_1/(2\epsilon^2)$ as $\epsilon \rightarrow
0$ and $m_{22}^*/(2m_{2}^*)=\Gamma$ (see \eqref{Gamma}).
Depending on the ratio $\gamma_2/\gamma_1$, $\Gamma$ and thus
$m_{22}^*$ are either positive or negative. The same phenomenon is
observed for bright solitons in systems of Bose-Einstein condensates
in optical lattices for which \eqref{Ham1} may be seen as a
tight-binding limit (see \cite{Mar03} and references therein).

\section{The $n=6$ case}
\begin{figure}
\includegraphics[scale=0.47]{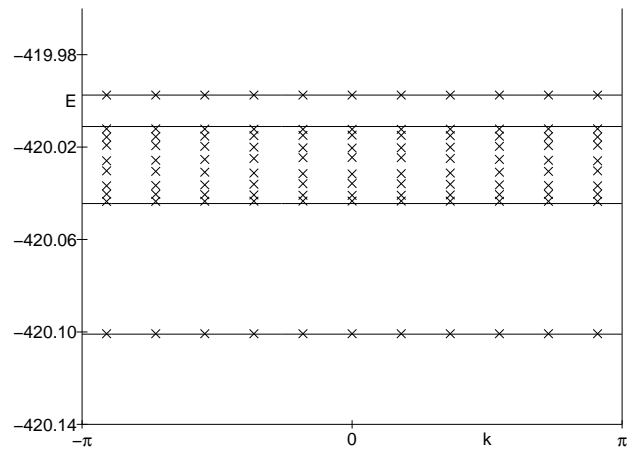}
\caption{\label{fig3} Detail of the eigenvalue spectrum for the generalized
  QDNLS periodic lattice (\ref{Ham2}), $n=6$, $f=11$, $\gamma_1=30,
  \gamma_2=0, \epsilon=0.5$.}
\end{figure}

As a further example we consider the $n=6$ case.  Now there are two bands
describing an interaction of two anharmonic states: the $\{4,2\}$ and
the $\{3,3\}$ bands.  Fig \ref{fig3}  shows a  $n=6$,  $\{4,2\}$
example, again the crosses represent numerically exact solutions in
the $f=11$ case, and the lines represent perturbation theory calculations.
In this case we have a ``continuum band'' which shows very weak
$k$-dependence, plus two ``line bands'', one above and one below.
With other choices of $\gamma_1$ and $\gamma_2$ we can move one or
both of the ``line bands'' into the continuum.

\begin{figure}
\includegraphics[scale=0.47]{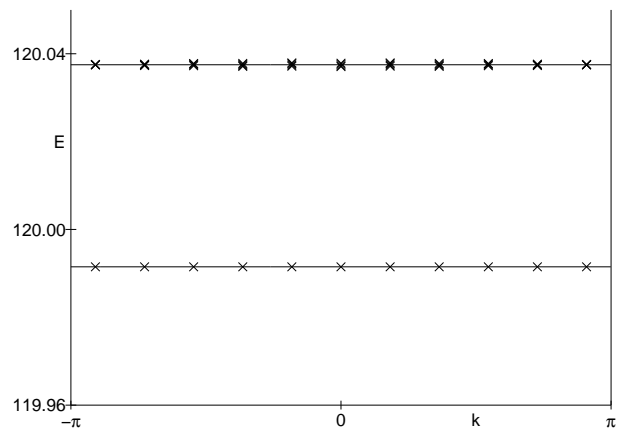}
\caption{\label{fig4} Detail of the eigenvalue spectrum for 
(\ref{Ham2}), $n=6$, $f=11$,
  $\gamma_1=10$, $\gamma_2=20$, $\epsilon=0.5$.}
\end{figure}

Fig \ref{fig4}  shows the corresponding $\{3,3\}$ case.
In this case there appears to be only two  ``line bands'', but a
closer examination reveals the upper ``line'' is in fact a continuum
band with the degeneracy split at an $O(\epsilon/\gamma)^3$ level.

\subsection{Perturbation theory, $n=6$, $\{4,2\}$ band}
%We proceed as in the $n=6$, $\{2,2\}$ case, but now states like
%$\ket{4002}$ and $\ket{2004}$ are distinct so there are more states to
%enumerate.  In 2nd order perturbation theory, the correction to the
%zeroth order calculation is given by the matrix
\noindent
Proceeding as in the $\{2,2\}$ case, we now obtain
\begin{equation} \label{H42}
H^{(42)} = D \epsilon^2\, I_{2\sigma} -\frac{\epsilon^2}{\gamma_1}
\left( 
\begin{array}{ccccc}
\Gamma & 1 &  &  & p \\
1 & 0 &  1 &  & \\
 & \ddots & \ddots & \ddots & \\
 &  &  1 & 0  & 1 \\
p^* &  &  & 1  & \Gamma \\
\end{array}
\right) 
\end{equation}   
where 
\begin{align*} 
D &=
-\frac{2}{3}\frac{5\gamma_1-9\gamma_2}{\gamma_1(\gamma_1-3\gamma_2)}\
;\ 
\Gamma = 
\frac{2}{3}\frac{4\gamma_1^2-27\gamma_2^2}
{(\gamma_1-3\gamma_2)(\gamma_1-6\gamma_2)};\\   
p &= \frac{6\gamma_1 e^{ik}}{\gamma_1-6\gamma_2} \, .
\end{align*}
An analysis of the eigenvalues of this matrix in the $f\rightarrow
\infty$ limit is somewhat messy but explicit results can be found:
details will be published elsewhere.  Essentially the results depend
on whether two rational functions of $\gamma_1/\gamma_2$ are less than
or greater than 1 in modulus.  This gives the two ``line bands'' in Fig
\ref{fig3}.  The corresponding eigenstates are
essentially symmetric and antisymmetric combinations of the Bloch
waves made from $\ket{42}$ and $\ket{24}$.

The ``continuum band'' is given by the formula
\begin{equation} \label{Ec42}
{E}_{c} =
24\gamma_2-14\gamma_1+\frac{2\epsilon^2}{\gamma_1}
\left[\frac{5\gamma_1-9\gamma_2}
{9\gamma_2-3\gamma_1}-\cos \theta\right]+{\cal 
  O}\left({\epsilon}^3\right),
\end{equation}
where $\theta \in (0,\pi)$.  Note that, up to order 2 in $\epsilon$,
the energies of this band do not depend on the crystal momentum $k$,
in accord with the numerical results.

\subsection{Perturbation theory, $n=6$, $\{3,3\}$ band}
In this case the correction to the zeroth order states is given by
\begin{equation} \label{H33}
H^{(33)} = \frac{6\epsilon^2}{3\gamma_2-2\gamma_1}\left( \, I_{\sigma} + 
M \right)
\end{equation}   
where $M_{1,1} = \Gamma=\frac{9}{2}(2\gamma_2-\gamma_1)/(\gamma_1-6\gamma_2)$
and $M_{i,j}=0$ otherwise.  The diagonal matrix (\ref{H33}) has an
``impurity'' $\Gamma$ responsible for a splitting of the states
$\ket{33}$ from the rest of the band. In this respect, the
$\ket{33}$'s can be seen as ``bound states of doublets'' within the
$\{3,3\}$-band.  Note again that none of the elements of the matrix
above contains the wave vector $k$. At this order of perturbation
theory, the bands are then flat. Moreover, the degeneracy of the
$\ket{3 \cdots 3}$'s (i.e. two 3-quanta breathers separated by one or
more empty sites) has still not been lifted although it would to next
order.
 
\section{Discussion}

It is possible to generalize the above results to discuss $\{m,\ell\}$
bands when $m+\ell=n>6$.  The case $m=\ell$ is similar to the
$\{3,3\}$ case and the case $m>\ell>1$ follows the $\{4,2\}$ case
discussed above.  Details will be given elsewhere.

The results known previously for bound states representing a group of
$n$ bosons located on the same site \cite{seg94} have been extended
here to higher order states representing two interacting groups of
bosons. This opens up the possibity of a study of the collision
process between these composite particles, that is a quantum breather
collision. A similar study has already been done for the {\em
  continuous} version of \eqref{Ham1}, i.e. the {\em integrable}
quantum nonlinear Schr\"odinger equation (QNSE), well known in
nonlinear optics \cite{Lai89} and quantum field theory
\cite{Thacker81}. However, the bound states of groups of quanta which
we have described here within the $\{m,\ell\}$ bands do not exist in
QNSE.  Their appearence in these {\em nonintegrable} discrete systems
is thus expected to affect the collision process between two quantum
breathers and to bring new features in comparison to the quantum
soliton collisions described in \cite{Lai89}.  Results of this
investigation will be reported elsewhere.

\begin{acknowledgments}
The authors are grateful for support under the LOCNET EU
network HPRN-CT-1999-00163.  This work was begun whilst JCE was
visiting Cambridge, and he thanks the Isaac Newton Institute
and Corpus Christi College for support.  JD and JCE would also like to
thank O.~Penrose for many useful discussions. 
\end{acknowledgments}
%\bibliographystyle{apsrev}
%\bibliography{/home/chris/latex/bibs/lattice}

\end{document}